\documentclass[%
 preprint,
 superscriptaddress,
 amsmath,amssymb,
 aps,
 showkeys,
 titlepage,
 endfloats*.   
]{revtex4-2}

\usepackage[utf8]{inputenc}
\usepackage[english]{babel}

\usepackage{graphicx}
\usepackage{dcolumn}
\usepackage{bm}

\usepackage[colorlinks = true,
            linkcolor = blue,
            urlcolor  = blue,
            citecolor = red,
            anchorcolor = blue]{hyperref}

\begin{document}


\title{Fast nanothermometry based on direct electron detection of electron backscattering diffraction patterns} 

\author{Ryan Gnabasik}
\affiliation{Department of Mechanical Engineering, University of California, Santa Barbara, CA 93106, USA}

\author{Razan O. Nughays}
\affiliation{Department of Mechanical Engineering, University of California, Santa Barbara, CA 93106, USA}
\affiliation{Center of Excellence for Renewable Energy and Storage Technologies, Division of Physical Science and Engineering, King Abdullah University of Science and Technology (KAUST), Thuwal 23955-6900, Kingdom of Saudi Arabia}

\author{Ashlynn Overholser}
\affiliation{Department of Mechanical Engineering, University of California, Santa Barbara, CA 93106, USA}

\author{Vijay Kumar}
\affiliation{Department of Mechanical Engineering, University of California, Santa Barbara, CA 93106, USA}

\author{Shantal Adajian}
\affiliation{Department of Mechanical Engineering, University of California, Santa Barbara, CA 93106, USA}

\author{Nicol\`{o} Maria della Ventura}
\affiliation{Materials Department, University of California, Santa Barbara, CA 93106, USA}

\author{Daniel S. Gianola}
\affiliation{Materials Department, University of California, Santa Barbara, CA 93106, USA}

\author{Bolin Liao}
\email{bliao@ucsb.edu} \affiliation{Department of Mechanical Engineering, University of California, Santa Barbara, CA 93106, USA}

\date{\today}

\begin{abstract}

Accurate temperature measurement at the nanoscale is crucial for thermal management in next-generation microelectronic devices. Existing optical and scanning-probe thermometry techniques face limitations in spatial resolution, accuracy, or invasiveness. In this work, we demonstrate a fast and non-contact nanothermometry method based on temperature-induced changes in electron backscattering diffraction (EBSD) patterns captured by a high-performance direct electron detector within a scanning electron microscope (SEM). Using dynamical electron simulations, we establish the theoretical temperature sensitivity limits for several semiconductors (Si, Ge, GaAs, and GaN), showing that thermal diffuse scattering (TDS) leads to a measurable smearing of Kikuchi bands in the EBSD patterns. We develop a Fourier analysis method that captures these subtle changes across the full diffraction pattern, achieving a simulated temperature sensitivity of approximately 0.15\% per K. Experimental results on silicon confirm a sensitivity of 0.14\% per K and achieve a 13-K temperature uncertainty with a 10-second acquisition time, and enable spatial temperature mapping under thermal gradients. Our approach offers a pathway toward practical and high-resolution thermal mapping directly in SEMs, expanding the toolbox for device-level thermal diagnostics.

\end{abstract}

                            
\maketitle


\section{Introduction}

Thermal management has become a pressing challenge in developing next generation microelectronic devices with rapidly increasing integration density and power consumption~\cite{qin2023thermal}. For example, peak heat fluxes in state-of-the-art power electronic devices can exceed those on the surface of the Sun. In addition, the rise of artificial intelligence is projected to drastically increase the power consumption in data centers that will require efficient thermal management strategies from the device level to the system level~\cite{li2024comprehensive}. These developments have called for a paradigm shift towards electro-thermal codesign of future electronic devices~\cite{choi2021perspective}. To enable further advancement in this direction, thermal diagnostic tools with high spatial resolution and temperature sensitivity are indispensible. 

At the device level, several major thermal issues need to be experimentally diagnosed and theoretically understood: local extreme hot spots in the conducting channel~\cite{schleeh2015phonon}, device/substrate interfacial heat conduction~\cite{malakoutian2025lossless}, and nonideal heat conduction near dislocations and other defects~\cite{mion2006accurate}. These issues critically determine the design strategies for the thermal management of microelectronic devices at the materials and system level. To experimentally probe these effects, however, a common obstacle is the associated small length scale on the nanometer (hot spots) and atomic scale (interfaces and defects), where thermal measurement becomes exceedingly difficult. Furthermore, noncontact thermometry methods are desirable to minimize the impact on device operation during measurements.

Existing optical thermometry methods, such as thermo-reflectance measurements, are intrinsically diffraction-limited to a spatial resolution above hundreds of nanometers when using visible light~\cite{jiang2018tutorial,schmidt2009frequency,choudhry2021characterizing}. Steady-state optical thermometry techniques, such as Raman spectroscopy~\cite{beechem2008micro}, can further suffer from the nonequilibrium distributions among different phonon groups~\cite{sokalski2022effects} and the complication caused by thermal stresses and electric fields~\cite{bagnall2017simultaneous} in addition to the limitation of their spatial resolution. Although super-resolution optical thermometry based on upconverting nanoparticles has been demonstrated to probe length scales below the diffraction limit~\cite{ye2024optical}, it requires the deposition of nanoparticles onto the sample as temperature sensors, which can create practical challenges when diagnosing electronic devices. Another category of thermometry techniques with nanometer spatial resolution relies on scanning probes. One example is scanning thermal microscopy (SThM)~\cite{zhang2020review}, where temperature sensors and/or heaters are fabricated onto nanoscale scanning tips. Despite the demonstrated nanometer spatial resolution and milli-Kelvin-level temperature sensitivity~\cite{reihani2024cooling}, the accuracy of SThM requires careful calibration and modeling of tip-sample interactions to avoid the interference of contact thermal resistance and topographical defects~\cite{menges2016nanoscale}. Another example is thermometry based on scanning near-field optical microscopy (SNOM), which can reach a spatial resolution down to tens of nanometers by combining an optical probe with a scanning tip or fiber~\cite{goodson1997near,ezugwu2017contactless}. However, parasitic heat transfer between the sample and the tip or fiber still poses challenges in interpreting these measurements.

Electron beams are alternative probes to realize nanometer-scale thermometry. Inside transmission electron microscopes (TEMs), multiple electron detection modes have been explored for nanoscale temperature mapping. For example, Wehmeyer et al. examined the temperature-dependent thermal diffuse scattering (TDS)~\cite{wang2003thermal} of electrons as a local temperature probe~\cite{wehmeyer2018measuring} and reported 5-K temperature uncertainty with an acquitision time of 96 s. TDS arises because of electron-phonon interactions that scatter electrons from their elastic Bragg reflections into the diffuse background. TDS has a higher temperature sensitivity than thermal expansion~\cite{niekiel2017local} and is universal in solid materials. Alternatively, electron energy loss spectroscopy (EELS) has been used to probe local temperature with nanometer spatial resolution based on the temperature-dependence of plasmon~\cite{mecklenburg2015nanoscale} or phonon peaks~\cite{idrobo2018temperature}. In general, TEM-based techniques require specialized fabrication of thin samples and are thus not suitable for thermal diagnosis of operating microelectronic devices. In contrast, scanning electron microscopes (SEMs) are more suitable platforms to probe microelectronic devices due to their minimal sample preparation requirements. Khan et al. have quantified the temperature dependence of secondary electron emission as a potential signal channel to map local temperatures inside an SEM~\cite{khan2018temperature} and reported an 8-K temperature resolution with a requisition time of 60 s. Alternatively, Wu and Hull demonstrated the temperature sensitivity of another detection mode in an SEM, electron backscatter diffraction (EBSD), where TDS disrupts the coherent electron diffraction process and thus reduces the intensity of the coherent Kikuchi bands while increasing the intensity of the diffuse background in EBSD diffraction patterns~\cite{wu2012novel}. With a conventional charge-coupled-device (CCD) EBSD detector, they realized a practical temperature resolution of around 10 K (with unspecified acquisition time) by analyzing the intensity of a single Kikuchi band in silicon (Si), which was limited by both inherent material properties and experimental conditions. Subsequently, they showed that the temperature dependence of the Debye-Waller $B$ factor, the backscatter electron yield, and the lattice constant are the main factors that control the inherent temperature sensitivity of a given material~\cite{wu2013material}. 

In this work, we further advance the EBSD-based nanothermometry in an SEM by (1) establishing the inherent temperature sensitivity in several technologically important semiconductors -- Si, germanium (Ge), gallium arsenide (GaAs), and gallium nitride (GaN) -- via dynamical electron simulation; (2) developing a new data processing protocol utilizing the smearing effect of Kikuchi bands in full EBSD patterns instead of an individual Kikuchi band; and (3) Obtaining significantly higher temperature resolution with shorter acquisition time using a direct electron detector. We further demonstrate spatial mapping of a temperature gradient in Si using this technique and highlight the remaining challenges. Our work advances the fundamental understanding of the temperature effect on EBSD patterns and paves the way towards practical nanothermometry within an SEM platform.

\begin{figure}[htp]
    \centering
    \includegraphics[width = \textwidth]{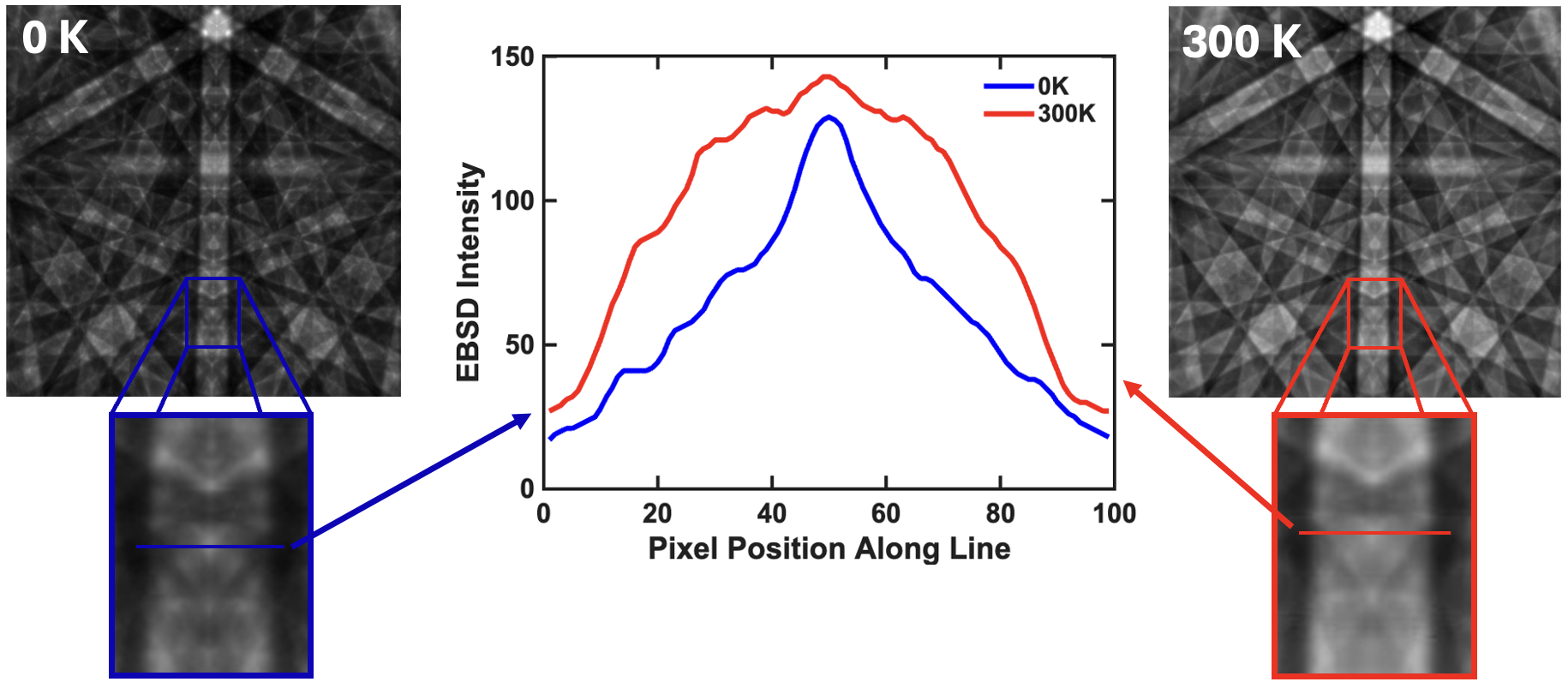}
    \caption{\textbf{Illustration of the effects of temperature on EBSD (Electron Backscatter Diffraction) patterns}. As the sample temperature increases, thermal diffuse scattering (TDS) intensifies, causing increased incoherent scattering of the electrons. This scattering leads to brighter and more diffuse EBSD patterns, as shown in the simulated EBSD patterns of silicon at 0 K and 300 K. The EBSD pattern at 0 K displays sharp features, while the 300 K pattern appears brighter with softened details due to increased TDS. The central plot compares the pixel intensity along a selected line in both images. While overall intensity increases at 300 K, the features become broader and less distinct. The blue curve (0 K) shows sharp intensity peaks that are still present but less defined in the 300 K case.}
    \label{fig:tds}
\end{figure}

\section{Methods}

\subsection{Theory and Computational Methods}
EBSD produces Kikuchi diffraction patterns in a two-step process~\cite{zaefferer2007formation}. First, the incident primary electrons hit the sample and experience incoherent scatterings with a wide range of scattering angles. The geometry of a typical EBSD setup (the sample is tilted at $\sim$70 degrees down from the horizontal plane and the EBSD detector is tilted $\sim$10 degrees away from the vertical plane) maximizes the signal of backscattered electrons. Second, the backscattered electrons interact elastically and coherently with the crystal lattice as they escape the sample. As a result, the Kikuchi patterns can be thought of as formed by electrons emitted from a small volume inside the sample that are subsequently coherently diffracted by the crystal lattice. Therefore, the geometry of the Kikuchi patterns obtained by EBSD has been widely used to identify crystallographic phases and orientations~\cite{humphreys2001review}. 

Temperature can affect the EBSD patterns in several ways. First, temperature changes can lead to thermal expansion or contraction of the lattice, which can affect the width of the Kikuchi bands. However, given the small thermal expansion coefficient of most crystalline solids on the order of $10^{-6}$ per K, the resulting temperature sensitivity given the spatial resolution of existing EBSD detectors is exceedingly low. Second, temperature changes can affect TDS that can in turn cause changes in both steps of the EBSD pattern formation. During the incoherent scattering step, TDS can increase the fraction of backscattered electrons at elevated temperatures~\cite{zaefferer2007formation}, which will be reflected in an increased overall intensity of the EBSD patterns, as observed by Wu and Hull~\cite{wu2012novel}. Nevertheless, the overall intensity of EBSD patterns is highly sensitive to experimental conditions and nonidealities, such as beam current fluctuation, sample surface condition, and detector position and alignment, and is thus difficult to calibrate and not ideal for thermometry. Besides the overall intensity of the EBSD patterns, TDS also reduces the coherent interaction of the electrons with the crystal lattice in the second step of EBSD pattern formation and thus shifts the contrast from the coherent Kikuchi bands to the diffuse background~\cite{zaefferer2007formation}, leading to a smearing of the Kikuchi bands as illustrated in Fig.~\ref{fig:tds}. Once the EBSD patterns are normalized by their average intensities, this contrast mechanism is robust against factors that can lead to fluctuations of the overall pattern intensity. As a first-order approximation, Wu and Hull discussed this TDS contrast using the Debye-Waller factor~\cite{wu2012novel,wu2013material}:
\begin{equation}
    \frac{I}{I_0}=\exp{(-2Bs^2)},
    \label{eqn:DWF}
\end{equation}
where $I$ is the intensity of a Kikuchi band with TDS, $I_0$ is the intensity of a Kikuchi band without TDS, $B$ is the Debye-Waller $B$-factor, and $s=\frac{\sin\theta}{\lambda}$, where $\theta$ is the Bragg diffraction angle and $\lambda$ is the wavelength of the electrons. The Debye-Waller $B$-factor is proportional to the mean squared displacement of atoms in a crystal lattice due to thermal vibrations and is thus responsible for the reduced intensity of coherent diffraction peaks or bands caused by TDS~\cite{wang2003thermal}. Although Eq.~\ref{eqn:DWF} provides an intuitive explanation of the TDS-induced reduction of the intensity of Kikuchi bands, it remains qualitative and cannot predict the local changes of entire EBSD patterns. Because of this limitation, the analysis by Wu and Hull~\cite{wu2012novel} only involved an individual Kikuchi band while a majority of information provided by the EBSD patterns was not fully utilized, thus limiting the achievable temperature sensitivity to about 0.018\% change in the Kikuchi band intensity per K.

In this work, we employ recently developed dynamical electron diffraction simulation to quantitatively analyze the impact of temperature-dependent TDS on EBSD patterns in technologically relevant semiconductors Si, Ge, GaAa, and GaN. Quantitative simulation of EBSD patterns has been a challenge due to the distinct characteristics of the two steps that form the patterns. In the first step of incoherent backscattering, typically Monte Carlo simulations are performed to track the trajectory of each incident primary electron and the BSEs~\cite{ouyang2023impact}. In the second step, dynamical electron diffraction of the BSEs on their way out of the sample needs to be simulated based on the Bloch wave formalism. Merging these two approaches into a unified framework is thus required to accurately simulate both the geometry and the intensity distribution of EBSD patterns while also incorporating the TDS effect based on the Debye-Waller factor~\cite{callahan2013dynamical}, as recently implemented in the open-source EBSD simulation software, EMSoft~\cite{singh2017emsoft,callahan2013dynamical}. Here, we use EMSoft to quantitatively simulate the temperature-dependent EBSD patterns of Si, Ge, GaAs, and GaN, which provides a refined theoretical limit of the achievable temperature sensitivity. For these simulations, the desired crystal is built with the material parameters such as the crystal structure, atom positions and occupations, and their respective Debye-Waller factor. Temperature-dependent lattice constants of the four materials are taken from experimental literature~\cite{okada1984precise,reeber1996thermal,vurgaftman2001band,roder2005temperature}. Debye-Waller factors were calculated using fit parameters from a density functional theory (DFT) simulation~\cite{schowalter2009computation}. From here, a Monte Carlo simulation is run with $2\times10^9$ incident primary electrons with 10 keV kinetic energy and a standard 70$^{\circ}$ sample tilt angle. The Monte Carlo simulation calculates the exit depth, energy, and direction of electrons that have been backscattered using the standard Bethe continuous-slowing-down approximation for charged particles entering crystalline materials. The information of these depth- and energy-dependent backscattered electrons is then used in a dynamical diffraction calculation based on the Bloch wave formalism to simulate an EBSD master pattern, which contains all of the diffraction space that can be sampled. Contributions from backscattered electrons up to 75 nm in depth were included for an energy range from 0 up to 10 keV. Once this master pattern is calculated, the relevant detector settings are applied and a gnomonic projection of the master pattern based on the reciprocal space sampled on the detector screen is then output as an EBSD pattern. Patterns were simulated to be 4000 pixels by 4000 pixels.

\subsection{Experimental Methods}

To further improve the temperature sensitivity at a given acquisition time, experimental EBSD patterns were taken on a custom direct electron sensing detector designed by Direct Electron LP (San Diego, CA USA) with optimum detection for electron beams between 8 to 12 keV with improved resolution, speed, and sensitivity~\cite{wang2021electron,della2025direct}. The detector model DE-SEMCam was integrated into a Thermo Fisher Apreo S SEM. This detector is a monolithic active pixel sensor (MAPS) that integrates the direct electron sensor and the readout electronics in a single wafer. It has a small pixel size of 13 $\mu$m and features a large array of pixels (2048 $\times$ 2048), which provides much improved angular resolution to detect high-index bands that are more sensitive to temperature~\cite{wu2012novel}. In addition, the DESEMCam detector showcases a high signal-to-noise ratio of 93 for operation at 10 kV~\cite{wang2021electron}. More details about the detector installation, operation, and performance were described in our previous papers~\cite{wang2021electron,della2025direct}.

    
To experimentally calibrate the temperature-dependent EBSD patterns of Si, we microfabricated platinum (Pt) wires on SiO$_2$/Si substrates to serve as both heaters and temperature sensors. The Pt microheaters were fabricated on a 1 cm $\times$ 2 cm silicon chip with a 1 $\mu$m silicon oxide (SiO$_2$) layer grown by dry thermal oxidation. 80-nm-thick Pt was deposited to form the heater and contact pads. The resistive element is 20 $\mu$m wide, and the contact pads measure 900 $\mu$m $\times$ 2 mm. To ensure the EBSD measurement is conducted directly on Si, inductively coupled plasma (ICP) etching was used to selectively etch SiO$_2$ in the regions between the resistors and contact pads. One resistive Pt element is used as the microheater and the adjacent (1 mm away) Pt element is used as a resistance temperature detector (RTD). For temperature sensing, the microfabricated RTD is calibrated by continuously measuring resistance change with temperature in an environmental chamber with precise temperature control. 

Experiments were performed at 10 keV with 6.4 nA beam current and an acquisition time of 10 s. Under these conditions, the steady-state temperature rise due to eletron beam heating is estimated to be less than 0.1 K~\cite{wu2012novel}. From the Monte Carlo simulation, most backscattered electrons escape from less than a 40-nm diameter sample surface area around the electron beam, defining the spatial resolution of our experiment. Patterns were collected (2048 by 2048 pixels) from regions of the sample close to the Pt wire being used as a resistive sensor. To acquire temperature-dependent EBSD patterns, the voltage was changed on the DC power supply and the sample would heat up via Joule heating. There was a minimum of 5 minutes between changes in voltage and capture of EBSD patterns in order for the temperature to stabilize and any adjustments for sample drift to be made. Patterns were taken several microns away from previous measurements to avoid beam-damage related effects. Small changes to focus or sample height relative to the electron beam and detector resulted in negligible changes to pattern sharpness and shift. Pt RTD measurements were taken after temperature stabilization and right before pattern acquisition.

\section{Results and Discussions}
\subsection{Simulation and Theoretical Limit of Temperature Sensitivity}
\begin{figure}
    \centering
    \includegraphics[width=1\linewidth]{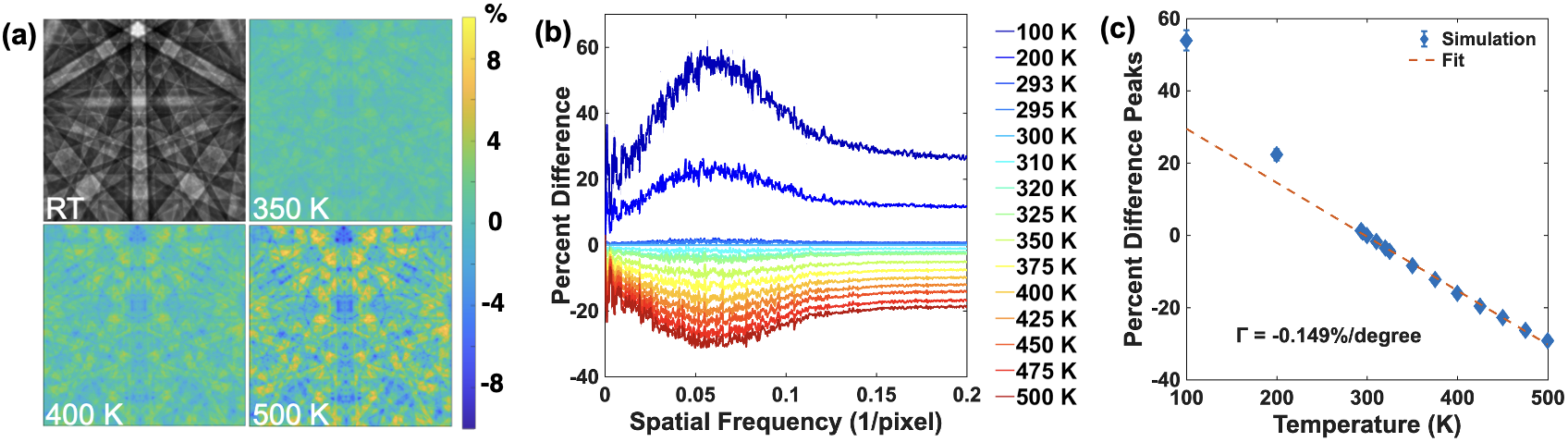}
    \caption{\textbf{Dynamical electron simulation of  temperature-dependent electron backscattering diffraction (EBSD) patterns in Si.} (a) Simulated EBSD pattern of Si at room temperature (293 K, RT) and the change in the EBSD patterns at 350 K, 400 K, and 500 K compared to that at RT, in percentage normalized to the RT intensity. (b) The radially integrated spatial Fourier transforms of EBSD patterns of Si simulated at different temperatures. Data at 293 K is used as a reference. Percentage change at other temperatures from the reference data are plotted. The noise in the data is due to the finite number of electrons used in the Monte Carlo step. (c) The peak change in the Fourier spectra (near the spatial frequency of 0.06 pixel$^{-1}$) as a function of temperature. The dashed line is a linear fit with a slope of -0.149\% per K. }
    \label{fig:simulation}
\end{figure}

\begin{figure}
    \centering
    \includegraphics[width=0.75\linewidth]{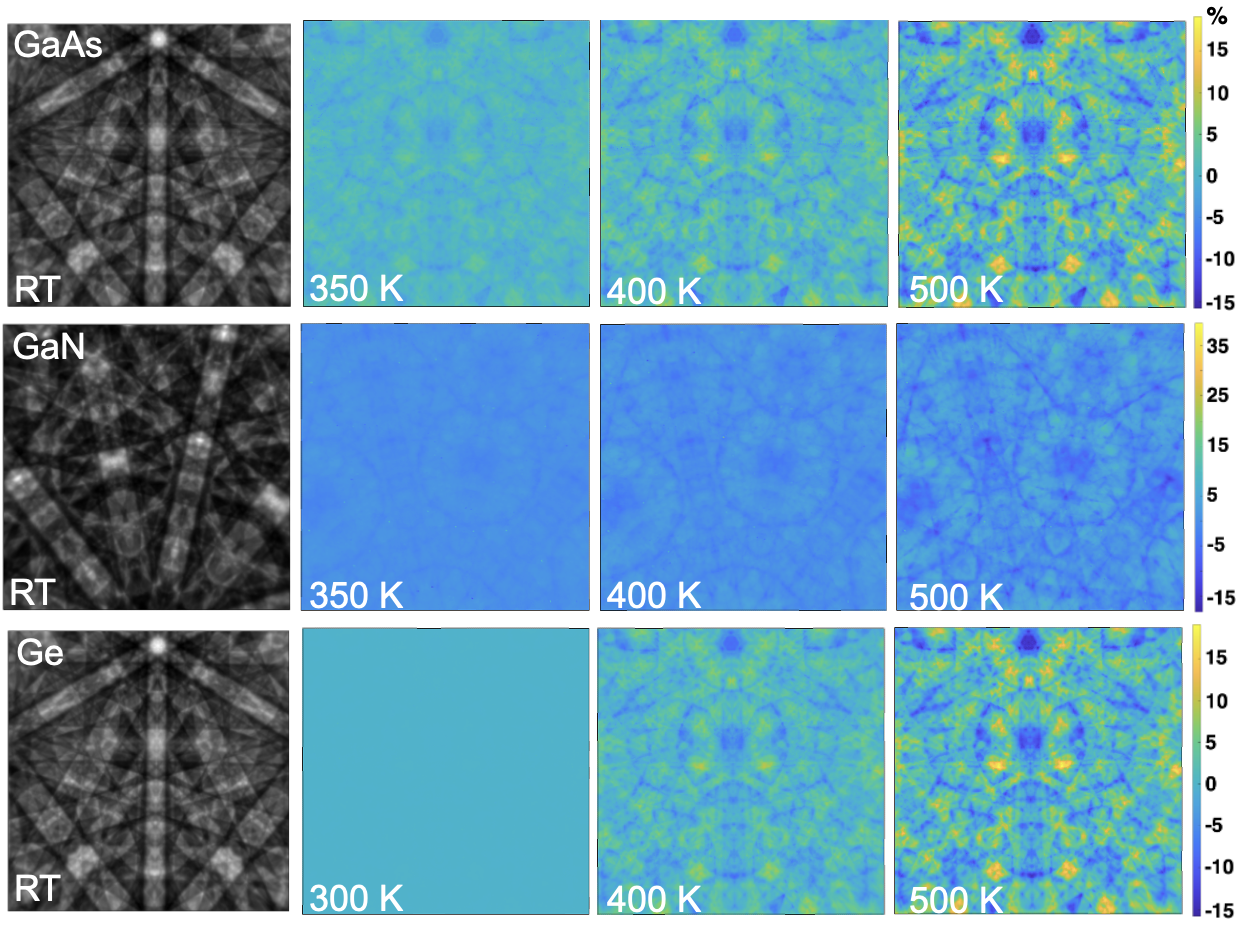}
    \caption{\textbf{Simulated temperature-dependent EBSD patterns of GaAs, GaN, and Ge.} The first column includes simulated EBSD patterns of GaAs, GaN, and Ge at room temperature (RT, 293 K), while the other columns show the change in the EBSD patterns at different temperatures compared to the RT patterns, in percentage normalized to the RT intensity. GaAs and Ge show similar changes in their EBSD patterns due to their identical crystal structures and similar atomic masses.}
    \label{fig:other-ebsd}
\end{figure}

We first examine the theoretical limit of the temperature sensitivity of EBSD patterns due to TDS. For this purpose, we conducted EMSoft simulations of temperature-dependent EBSD patterns for Si, Ge, GaAs, and GaN. The temperature effects are incorporated into the simulation by taking into account thermal expansion through temperature-dependent lattice constants and TDS through the Debye-Waller factors, although the thermal expansion effect is expected to be negligible compared to TDS. Figure~\ref{fig:simulation}a shows the simulated EBSD patterns in Si, where the top left panel shows the simulated pattern at 293 K (room temperature, RT) as a reference while the other three panels show the change of the patterns at 350 K, 400 K, and 500 K from the reference. All EBSD patterns shown in this paper are normalized by their average intensity. Simulated patterns for Ge, GaAs, and GaN are provided in Fig.~\ref{fig:other-ebsd}. From the simulated normalized patterns, it can be observed that, in general, the intensity of the high-symmetry Kikuchi bands is reduced, while the intensity of the diffusive background increases as the temperature rises. In particular, the zone axes (band crossings) experience the most reduction in intensity, suggesting that near-zone (high-symmetry) orientations are more susceptible to TDS. This observation is consistent with the theoretical expectation that increased TDS will reduce the portion of electrons that experience coherent interactions with the lattice to form the Kikuchi bands. 

\begin{figure}
    \centering
    \includegraphics[width=0.5\linewidth]{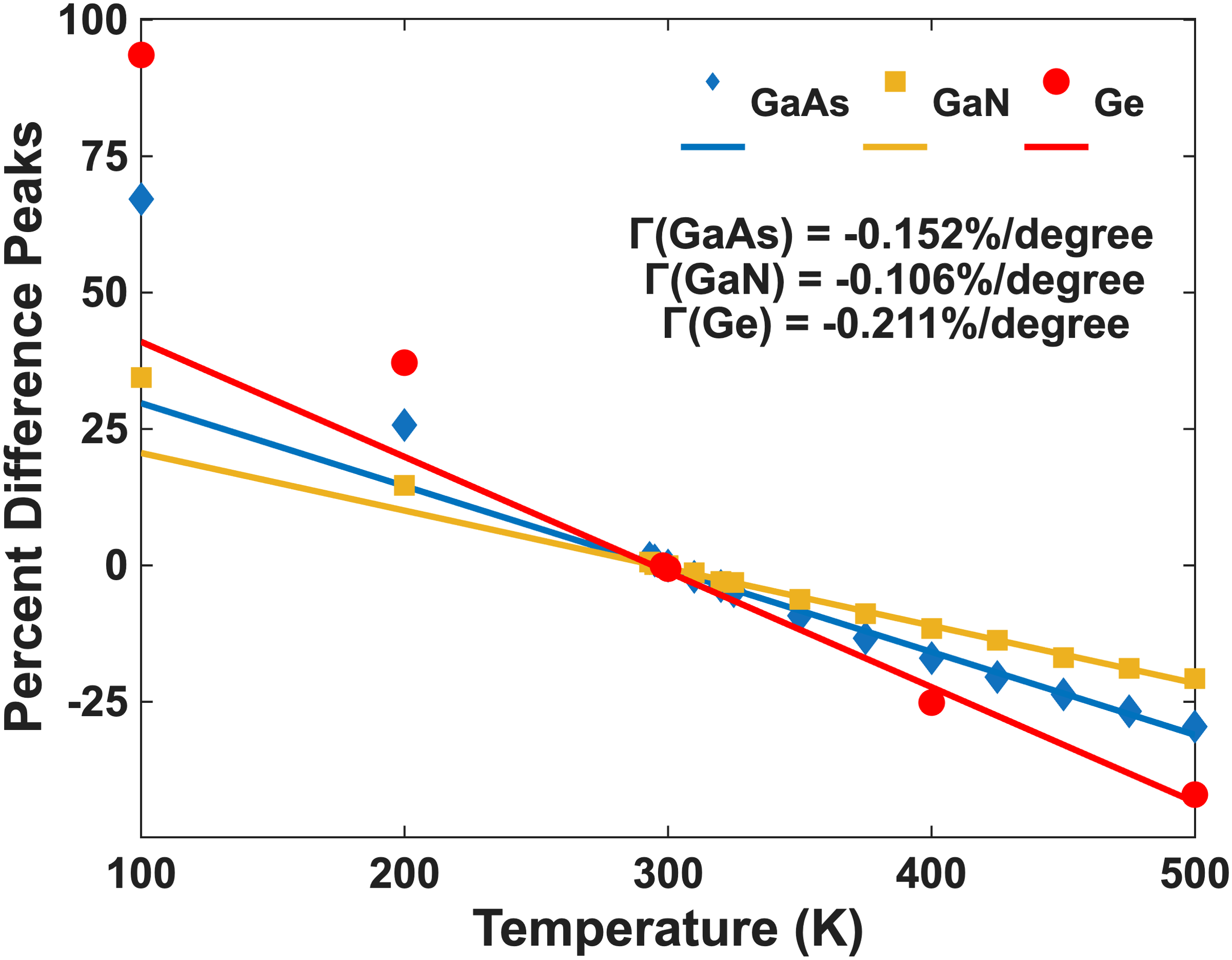}
    \caption{\textbf{Temperature coefficients for GaAs, GaN, and Ge.} The peak change in the Fourier spectra as a function of temperature for simulated EBSD patterns of GaAs, GaN and Ge. The linear fits are represented by the solid lines with the slope labeled in the figure. }
    \label{fig:tc}
\end{figure}

Given the observed changes in the EBSD patterns, one key question is how the patterns should be analyzed to maximize the obtainable temperature sensitivity. Wu and Hull chose to analyze the intensity distribution along a line cut across a high-index Kikuchi band, by which they obtained a temperature sensitivity of 0.018\% per K~\cite{wu2012novel}. However, this method did not make use of the majority of information contained in EBSD patterns. To address this issue, we propose a data processing method that utilizes the information from full EBSD patterns. The rationale behind our method is that increased TDS at elevated temperatures causes intensity transfer from coherent Kikuchi bands to the diffusive background, leading to a smearing effect of the Kikuchi bands, especially at their edges (Fig.~\ref{fig:tds}). Since the band edges are associated with high spatial frequencies of the EBSD patterns, we expect to see significant changes of the intensity at these spatial frequencies. To demonstrate this concept, we performed two-dimensional (2D) spatial Fourier transforms of the EBSD patterns and then conducted a radial integration of the intensity of the Fourier components. The results of the processing of the simulated Si patterns between 100 K and 500 K are shown in Fig.~\ref{fig:simulation}b, where the percentage change in intensity of each spatial frequency from its reference value at 300 K is shown. The noise in the data is due to the finite number of electrons used in the Monte Carlo step of the simulation. It is observed that the intensity associated with spatial frequency around 0.06 pixel$^{-1}$ shows the highest temperature sensitivity. In Fig.~\ref{fig:simulation}c, we plot the percentage change in intensity of this spatial frequency as a function of temperature, using its value at 300 K as the reference. Above 300 K, we found that the intensity changes almost linearly with temperature, showing a temperature sensitivity of 0.149\% per K, which is roughly one order of magnitude higher than that achieved by Wu and Hull and on par with the temperature sensitivity obtained by Wehmeyer et al. using TDS in TEM~\cite{wehmeyer2018measuring}. The main reason for this improvement is that the temperature effects on all Kikuchi bands in an EBSD pattern contribute to the obtained temperature sensitivity using our processing method. Similar analyses of Ge, GaAs, and GaN are provided in Fig.~\ref{fig:tc}, which shows that Ge has the highest temperature coefficient while GaN has the lowest. The difference in the temperature coefficients among different materials is mainly caused by the temperature-dependence of their Debye-Waller factors. 

\subsection{Experimental Results}
\begin{figure}
    \centering
    \includegraphics[width=0.75\linewidth]{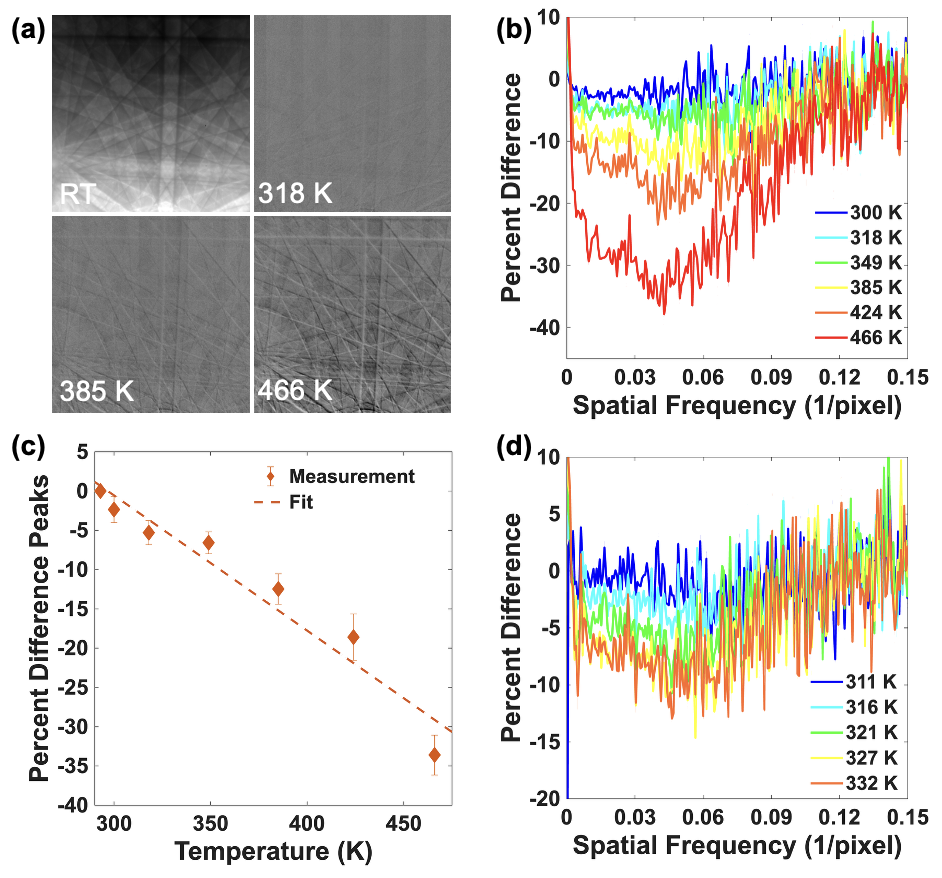}
    \caption{\textbf{Experimental measurement of temperature-dependent electron backscattering diffraction (EBSD) patterns in Si.} (a) Measured EBSD pattern of Si at room temperature (293 K, RT) and the change in the EBSD patterns at 318 K, 385 K, and 466 K compared to that at RT. (b) The radially integrated spatial Fourier transforms of EBSD patterns of Si measured at different temperatures. Data at 293 K is used as a reference. Percentage change at other temperatures from the reference data are plotted.  (c) The peak change in the Fourier spectra (near the spatial frequency of 0.04 pixel$^{-1}$) as a function of temperature. The error bars represent the standard deviation of the Fourier amplitude within the spatial frequency range of 0.03 and 0.05 pixel$^{-1}$. The dashed line is a linear fit with a slope of -0.14\% per K. (d) The radially integrated spatial Fourier transforms of EBSD patterns of Si measured at different temperatures with 5-K intervals.}
    \label{fig:experiment}
\end{figure}

Although the simulations provide the theoretical limit of the temperature sensitivity of EBSD patterns, the practically achievable temperature resolution depends strongly on the detector characteristics and the acquisition time. In particular, the acquisition time is an important parameter that determines the practicality of temperature mapping in realistic material structures and devices. For example, Wehmeyer et al.~\cite{wehmeyer2018measuring} achieved 5 K temperature sensitivity in gold films with an acquisition time of 96 s using TDS in STEM, which indicates that a temperature map containing hundreds (or more) pixels would be impractical. In Fig.~\ref{fig:experiment}, we present our experimental EBSD patterns taken from a Si sample in the temperature range of 293 K to 466 K. Figure~\ref{fig:experiment}a shows the experimental EBSD pattern taken at 293 K (RT) and the difference patterns at elevated temperatures. The experimental EBSD patterns at higher temperatures show similar changes as those observed in the simulated patterns in Fig.~\ref{fig:simulation}a, namely a reduced intensity in the coherent Kikuchi bands and an increased intensity in the diffusive background. These changes lead to the smearing of the Kikuchi band edges. To verify the temperature sensitivity obtained in the simulation, we first performed a set of measurements in a wider temperature range with coarser steps. Figure~\ref{fig:experiment}b shows the radially integrated Fourier transforms of the EBSD patterns taken at temperatures between 293 K and 466 K, where the data taken at 293 K was used as a reference and subtracted from those taken at higher temperatures. These EBSD patterns were taken with an acquisition time of 10 s. These experimental curves agree quantitatively with the simulation shown in Fig.~\ref{fig:simulation}b and the intensity associated with the Fourier component most sensitive to temperature shows a temperature sensitivity of roughly 0.14\% per K (as shown in Fig.~\ref{fig:experiment}c), which is slightly lower than the simulation result. In addition, we quantify the uncertainty of these measurements by calculating the standard deviation of the Fourier transform magnitude in the spatial frequency range of 0.03 to 0.05 pixel$^{-1}$, which is a measure of the noise level around the peak spatial frequency. The typical uncertainty is about 2\%, corresponding to a temperature uncertainty $u_T$ of about 13 K, which is larger than that achieved by Wehmeyer et al.~\cite{wehmeyer2018measuring} but with a much shorter acquisition time.
To further test the practical temperature resolution, we conducted another set of measurements with finer steps between 311 K and 337 K with 5 K intervals. The result is shown in Fig.~\ref{fig:experiment}d, where the data taken at 293 K was chosen as the reference. Despite the noise associated with each curve, the small temperature increments were clearly discernible. We emphasize here that these measurements were made with short acquisition times (10 s).

\begin{figure}
    \centering
    \includegraphics[width=1\linewidth]{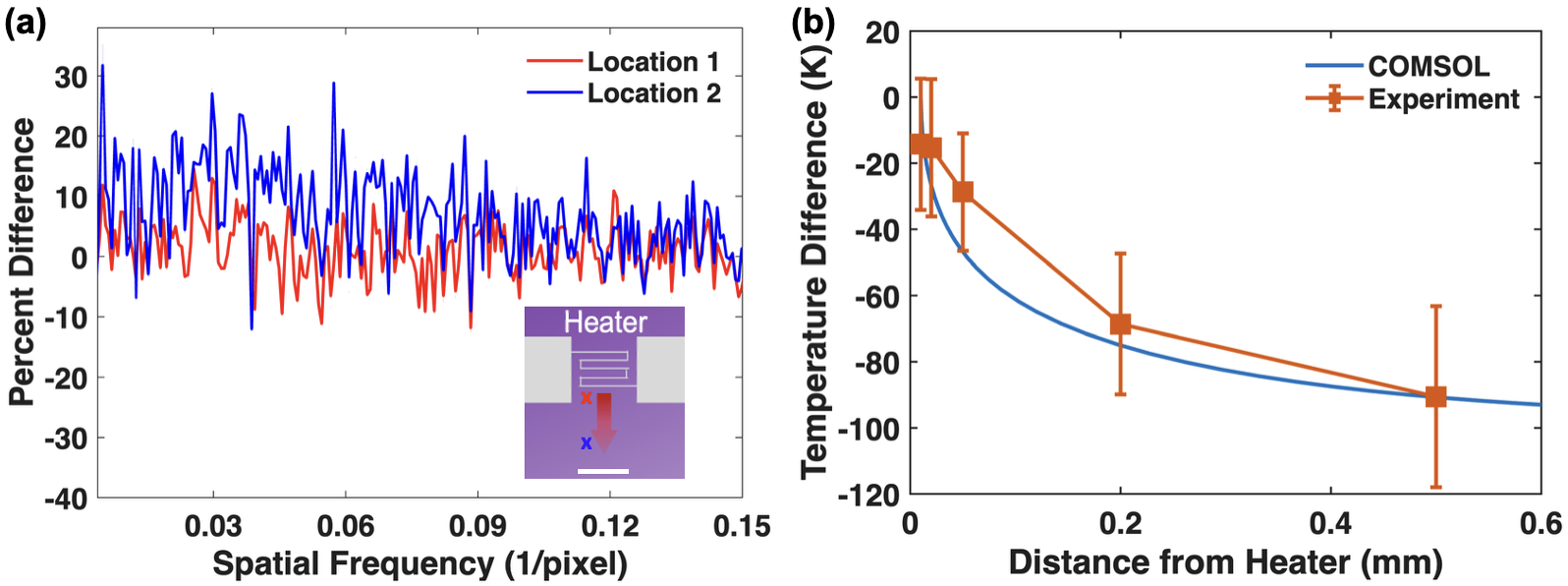}
    \caption{\textbf{Experimental mapping of a temperature gradient using electron backscattering diffraction (EBSD) patterns in Si.} (a) The radially integrated spatial Fourier transforms of EBSD patterns of Si measured at two different locations as marked in the inset after subtracting a reference data taken right at the edge of the heater. Location 1 is 10 $\mu$m away from the heater, while location 2 is 500 $\mu$m away from the heater. The inset shows an optical image of the fabricated Pt heater. The scale bar: 500 $\mu$m. (b) The measured temperature distribution as a function of distance from the edge of the heater. The error bars are calculated as the standard deviation of the Fourier transform intensity within the spatial frequency range of 0.03 and 0.05 pixel$^{-1}$. The blue solid line is from a finite element simulation using COMSOL.}
    \label{fig:mapping}
\end{figure}

We also explore spatial temperature mapping by acquiring EBSD patterns at different locations in Si with a temperature gradient, which was not demonstrated in previous works by Wehmeyer et al.~\cite{wehmeyer2018measuring} and Wu and Hull~\cite{wu2012novel}. As shown in the inset of Fig.~\ref{fig:mapping}a, a patterned Pt heater is used to create a temperature gradient. The temperature within the heater area is estimated to be around 393 K by measuring the resistance of the Pt heater. EBSD patterns were taken at selected locations up to 500 $\mu$m away from the edge of the heater. The measurement locations were selected to be denser closer to the heater, where the temperature gradient was the greatest. A reference EBSD pattern was taken right at the edge of the heater. For clarity, the Fourier transforms of the two EBSD patterns taken at 10 $\mu$m and 500 $\mu$m from the heater are shown in Fig.~\ref{fig:mapping}a, after subtracting the reference data. The data is significantly noisier, potentially due to larger temperature fluctuations associated with a significant temperature gradient (about 90 K temperature drop over 500 $\mu$m). Using the average Fourier transform magnitude within the spatial frequency range of 0.03 and 0.05 pixel$^{-1}$ and the experimentally determined temperature coefficient (0.14\% per K), the temperature distribution can be calculated and the result is shown in Fig.~\ref{fig:mapping}b. Together we show the expected temperature distribution obtained using a finite element simulation with COMSOL. It is clear from the result that spatial temperature mapping remains challenging even with the direct electron detector, especially at locations with a large temperature gradient, where our measurements appeared to overestimate the local temperature. One possible source of error could be the significant thermal stress associated with the large temperature gradient, which is known to impact the EBSD patterns~\cite{wright2011review,wang2023dislocation}. More detailed EBSD simulation and calibration are required to account for this effect. In addition, increasing the acquisition time at each location is one way to reduce the temperature uncertainty. However, prolonged beam exposure can lead to sample degradation that can introduce measurement artifacts (Wehmeyer et al.~\cite{wehmeyer2018measuring} circumvented this issue by accumulating counts over multiple sample locations with the same temperature, which is, however, not feasible in practical temperature mapping). Other data processing methods that make use of the full EBSD patterns, especially with the help of machine-learning-based techniques, can also potentially improve the accuracy for temperature mapping. Another potential route is to lower the primary electron energy, which can both enhance the temperature sensitivity and increase the spatial resolution. However, direct electron detectors optimized for these low-energy primary electrons need to be developed for this purpose.

\section{Conclusion}
In summary, we improved EBSD-based nanothermometry technique capable of achieving high temperature sensitivity and nanometer-scale spatial resolution using a direct electron detector in an SEM. Through dynamical electron simulations, we demonstrated that temperature-induced thermal diffuse scattering causes predictable smearing of Kikuchi bands, which we quantified using a full-pattern Fourier analysis. This method significantly improves the sensitivity over previous single-band analyses. Experimentally, we achieved a temperature sensitivity of ~0.14\% per K and a 13-K temperature uncertainty with a 10-second acquisition time, enabling practical application in nanoscale thermal mapping of microelectronic structures. While further improvements in detector optimization, data processing, and beam energy control could enhance resolution and accuracy, our approach opens a promising route for fast, non-invasive temperature measurements at the nanoscale. Future work may incorporate machine learning algorithms to further improve sensitivity and enable real-time temperature mapping in complex device environments.

\begin{acknowledgments}
We thank Alex Ackerman, McLean Echlin, and Marc De Graef for helpful discussions. We thank Kalani Moore in Direct Electron LP for help with the direct electron detector. This work is based on research supported by the Office of Navel Research under award number N00014-22-1-2262 and the Army Research Office under award number W911NF2310188. N.d.V. and D.S.G. acknowledge support from the Army Research Laboratory accomplished under Cooperative Agreement Number W911NF-22-2-0121. The scanning electron microscope used in this work is hosted in the UCSB Materials Research Laboratory Shared Experimental Facilities, which are supported by the MRSEC Program of the NSF under Award Number DMR-2308708. A portion of this work was performed in the UCSB Nanofabrication Facility, an open access laboratory. 
\end{acknowledgments}

\bibliography{references.bib}

\end{document}